\documentstyle[prb,aps,twocolumn,floats]{revtex}
\input psfig.sty
\newcommand{\imag}{i}
\begin{document}
\draft
\title{Magnetism of a tetrahedral cluster spin-chain}
\author{Wolfram Brenig}
\address{Institut f\"ur Theoretische Physik, Technische Universit\"at
Braunschweig, 38106 Braunschweig, Germany}
\author{Klaus W. Becker}
\address{Institut f\"ur Theoretische Physik, Technische Universit\"at
Dresden, 01062 Dresden, Germany\vskip.2cm}
\author{Dedicated to Professor Erwin M\"uller-Hartmann on the occasion
of his 60$^{\rm th}$ birthday.\vskip.3cm}

\date{\today}
\maketitle
\begin{abstract}
We discuss the magnetic properties of a dimerized and completely
frustrated tetrahedral spin-1/2 chain. Using a combination of exact
diagonalization and bond-operator theory the quantum phase diagram is shown
to incorporate a singlet-product, a dimer, and a Haldane phase. In
addition we consider one-, and two-triplet excitations in the dimer phase
and evaluate the magnetic Raman cross section which is found to be strongly
renormalized by the presence of a two-triplet bound state. The link to a
novel tellurate materials is clarified.
\end{abstract}


\section{Introduction}
\begin{figure}[bth]
\vskip.2cm
\centerline{\psfig{file=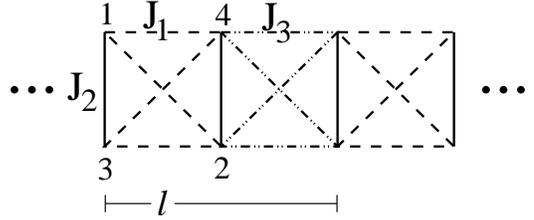,width=7cm}}
\vskip.2cm
\caption{The tetrahedral cluster-chain. $l$ labels the unit cell
containing spin-1/2 moments ${\bf s}_{i\,l}$ at the vertices
$i=1,\ldots,4$.} 
\label{fig3}
\end{figure}

\ Low-dimensional quantum-magnetism has received considerable interest
recently due to the discovery of numerous novel materials with
spin-$\frac{1}{2}$ moments arranged in chain, ladder, and depleted planar
structures. Many of these materials exhibit unconventional magnetic phases
due to dimerization and frustration of antiferromagnetic exchange.
Particular effort\cite{MH00} has been devoted to systems like
SrCu$_2$(BO$_3$)$_2$\cite{Kageyama99,Miyahara99}, which display a complete
frustration of the magnetic exchange as in the two-dimensional
Shastry-Sutherland model\cite{Shastry81}. In one dimension complete
frustration can occur in two-leg spin-ladders if an additional cross-wise
exchange is included as depicted in fig.~\ref{fig3} which resembles a
chain of edge sharing tetrahedra. For $J_1=J_3$ such tetrahedral ladders
have been investigated in the past\cite{Bose93,Ghosh97,Honecker00}. Very
recently, tellurates of type Cu$_2$Te$_2$O$_5$X$_2$ with X=Cl, Br have
been identified as a new class of spin-1/2 tetrahedral-cluster
compounds\cite{Johnsson00}. Bulk thermodynamic data have been analyzed in
the limit of isolated tetrahedra\cite{Johnsson00}. Raman spectroscopy,
however indicates a substantial inter-tetrahedral coupling along the
c-axis direction of Cu$_2$Te$_2$O$_5$X$_2$\cite{Lemmens01z}.  In this
direction the exchange topology is likely to be analogous to that of
fig.~\ref{fig3} with $J_1\neq J_3$. From a materials perspective it is an
open question if the magnetism of the tellurates can be understood in terms
of a dimerized tetrahedral spin-ladder.  From a theoretical point of view,
however, the magnetic properties of this model are an interesting issue
which forms the motivation for this work.

The paper is organized as follows. In the remainder of this section we
discuss the basic properties of the tetrahedral chain hamiltonian. In
section two the quantum phase diagram is analyzed. In section three a
bond-operator method is applied to the tetrahedral chain and in section 
four the magnetic Raman cross-section is evaluated.

The hamiltonian of the tetrahedral chain can be written in terms of the
total edge-spin operators ${\bf T}_{1(2)\,l}={\bf s}_{1(4)\,l}+{\bf
s}_{3(2)\,l}$ and the dimensionless couplings $b=J_3/J_1$ and $a=J_2/J_1$
\begin{eqnarray}
\frac{H}{J_1}=
\sum_l [{\bf T}_{1l}{\bf T}_{2l}
+b {\bf T}_{2l}{\bf T}_{1l+1}
+\frac{a}{2} ({\bf T}^2_{1l} +
{\bf T}^2_{2l})  -\frac{3a}{2}] 
\label{bb1}
\end{eqnarray}
This model displays infinitely many local conservation laws: $[H,{\bf
T}^2_{i(=1,2)\,l}]=0;\,\forall\,l,i=1,2$. Therefore, the Hilbert space
decomposes into sectors of fixed distributions of edge-spin eigenvalues
$T_{i\,l}=1$ or $0$, each corresponding to a sequences of spin-1
chain-segments intermitted by chain-segments of {\em localized} singlets.
If $J_1\neq J_3$ the spin-1 chain-segments are dimerized. In the infinite
length, dimerized $S=1$-chain sector, i.e.~for $T_{i\,l}=1$ $\forall i,l$,
the model simplifies to
\begin{eqnarray}
\frac{H}{J_1}=
\sum_l [{\bf S}_{1l}{\bf S}_{2l}
+b {\bf S}_{2l}{\bf S}_{1l+1}]
+\frac{a}{2}D 
\label{w3}
\end{eqnarray}
where ${\bf S}_{il}$ refer to spin-1 operators and $D$ is the number of
tetrahedra ('dimers').

\begin{table}[ht]
\begin{tabular}{c|c|c|c}
 & $T_1$ & $T_2$ & $E/J_1$ \\ \hline
${\cal S}_1$     & 1 & 1 & -2+a/2 \\
${\cal S}_2$     & 0 & 0 & -3a/2 \\
${\cal T}_1$     & 1 & 1 & -1+a/2 \\
${\cal T}_{2,3}$ & 0,1 & 1,0 & -a/2 \\
${\cal Q}$       & 1 & 1 & 1+a/2
\end{tabular}
\caption{Eigenstates and energies of the tetrahedron. Columns $T_1,2$
refer to corresponding edge-spin quantum number, site index $l$ suppressed.}
\label{tab0} 
\end{table}

The Hilbert space of a single tetrahedron consists of 16 states, i.e., two
singlets ${\cal S}_{1,2}$, three triplets ${\cal T}_{1,2,3}$ and one quintet
${\cal Q}$ the energies and $T_i$ quantum numbers of which are listed in
table~\ref{tab0}. Johnsson and collaborators\cite{Johnsson00} have first
pointed out that this level scheme implies a singlet to reside within the
singlet-triplet gap of the tetrahedron for $1/2<a<2$. Moreover the ground
state switches from ${\cal S}_1$ to ${\cal S}_2$ at $a=1$ suggesting a
line of quantum phase transitions in the $(a,b)$-plane for the lattice
model.

\section{Quantum Phase Diagram}
In this section we discuss the ground state of the tetrahedral chain. To
begin, we note that by a shift of one half of the unit cell, 
i.e.~${\bf T}_{2\,l(1\,l+1)}\rightarrow {\bf T}_{1\,l(2\,l)}$, model (\ref{bb1})
is symmetric under the operation $(J_1,a,b)\rightarrow (J_1 b,a/b,1/b)$.
Therefore, in order to cover the {\em complete} parameter space for $a,b>0$
it is sufficient to consider the phase diagram in the range of
$a\in[0,\infty]$ and $b\in[0,1]$.

Next we note, that the ground state of (\ref{bb1}) will be either in the
dimerized $S=1$-chain sector or in a homogeneous product of ${\cal S}_2$
states only. Inhomogeneous phases consisting of both, $T_{i\,l}=0$ {\em and}
$T_{i\,l}=1$ sites are not allowed for as ground-states. To see this we fix
$b$ and assume $a\rightarrow\infty$, in which case the ground state is a
pure product of ${\cal S}_2$-type singlets: $|\psi_0\rangle=\prod_l|\tilde{s}_l
\rangle$. Next we check for the ground-state energy change $\Delta E(a,b,N)$
upon forming a single connected chain-segment of length $N$ composed out of
$T_{i\,l}=1$-sites within the homogeneous state $|\psi_0\rangle$. To be
specific we first assume the chain-segment to consist of $D'=N/2$ tetrahedra
in which case
\begin{eqnarray} 
\Delta E(a,b,D')= D'[2(a-1)-e(b,D')] \;.
\label{qp1}
\end{eqnarray}  
Here $-e(b,D')<0$ is the ground-state energy gain {\em per two sites} due to
the inter-tetrahedral coupling. The main point is, that $e(b,D')$ is a
monotonously {\em increasing} function\cite{YesWeDidDeltaEofN} of $D'$.
Therefore the largest critical value $a_c=max\{a_c(D')\}$ at which the
formation of tetrahedra in the $S=1$ sector is favorable, i.e.~at which
$\Delta E(a_c(D'),b,D')$ turns negative, results for $D'\rightarrow\infty$.
This implies a single first order quantum phase transition into the
infinite-length,
dimerized $S=1$-chain sector as a function of decreasing $a$. Similar
arguments can be pursued for odd $N$.

\begin{figure}[tb]
\vskip0cm
\centerline{\psfig{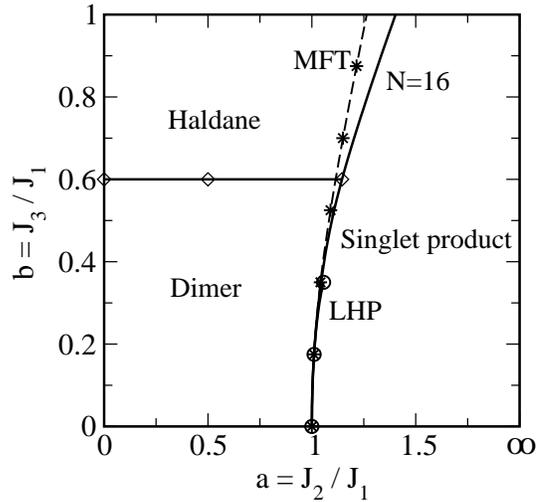}}
\vskip.5cm
\caption{Quantum phase diagram of the tetrahedral chain. Bare solid line:
1st-oder transition from ED for N=16 sites and PBC at 41 values of
$b\in[0,1]$. The critical value $a_c$ at $b=1$ from ED is $a^{N=16}_c(b=1)
\simeq 1.40292$. Solid line with diamond markers: 2nd-order Haldane-Dimer
transition  at $b\simeq 3/5$, extrapolated from ED (see fig.~\ref{qp3} and
ref.$^{14,15}$). Dashes(Solid) line with stared(circled)
markers refers to bond-boson mean-field/MFT (Holstein-Primakoff/LHP)
approach. LHP terminates at $b=3/8$.}
\label{qp2}                                      
\end{figure}

In fig.~\ref{qp2} we show the quantum phase diagram. From (\ref{qp1}) the
first order critical line $a_c(b)$ between the infinite-length, dimerized
spin-1 chain for $a<a_c(b)$ and the ${\cal S}_2$-type singlet
product-state for $a>a_c(b)$ is fixed by $a_c(b)=1+e(b)/2$, where
$e(b)=\lim_{D'\rightarrow \infty}e(b,D')$. To determine $e(b)$ we have
calculated the ground-state energy of dimerized spin-1 chains using exact
diagonalization (ED) with periodic boundary conditions (PBC) on up to
$N=16$ sites and a bond-boson theory the results of which will be detailed
in section~\ref{sec3}.  Regarding the ED the critical value of
$a_c(b=0)=1$ agrees with ref.\cite{Johnsson00}, while $a^{N=16}_c(b=1)\simeq
1.403$ agrees with ref.\cite{Sakai91} and is consistent with an extrapolated
value of $a^{N=\infty}_c(b=1)\simeq 1.401$ from
Density-Matrix-Renormalization-Group (DMRG)
calculations\cite{White93,Honecker00} and ED on 22 sites\cite{Golinelli94}.

\begin{figure}[tb]
\vskip0cm
\centerline{\psfig{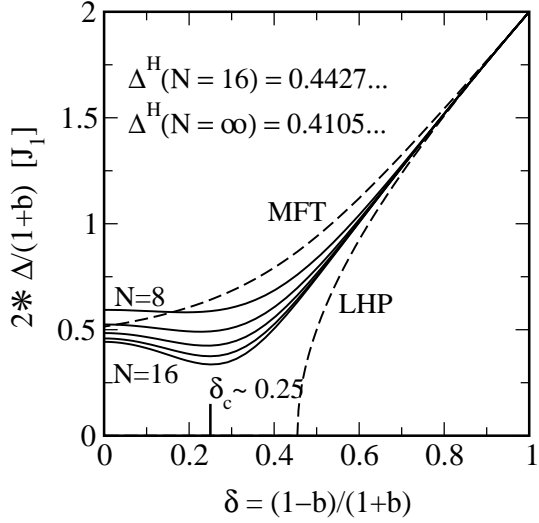}}
\vskip.5cm
\caption{Solid lines: spin gap $\Delta$ from ED for $N=8$, $10$, $12$,
$14$, and $16$
sites and PBC in the dimerized spin-1 chain sector at 41 values of the
inter-tetrahedral coupling $b\in[0,1]$.  Axis have been scaled to allow
for a comparison with ref.$^{14}$.  $\Delta^H$ refers to the spin
gap at $b=1$, i.e. the Haldane gap.  $\Delta^H(N=16)$ as from this work
and $\Delta^H(N=\infty)$ as from ref.$^{12}$.  Upper(lower) dashed
line: spin gap from bond-boson mean-field/MFT (Holstein-Primakoff/LHP)
approach.}
\label{qp3}          
\end{figure}

Within the dimerized $S=1$-chain sector an additional second-order quantum
phase transition exists between the dimer phase for $b<b_c$ and the
Haldane phase for $b>b_c$. This transition has been studied extensively
(see eg.\cite{Totsuka95} and refs.~therein), resulting in $b_c\simeq 3/5$
from DMRG calculations\cite{Kato94} and finite-size scaling
analysis\cite{Totsuka95}. However, this transition is not at the focus of our
study. In fig.~\ref{qp3} our ED results on the
finite-size behavior of the spin gap in the dimerized $S=1$-chain sector
are shown as function of $b$, which signals the dimer-Haldane transition and
directly reproduces identical data which have been obtained 
earlier by Kato and Tanaka\cite{Kato94}. Fig.~\ref{qp3} contains 
additional results for the spin gap from the bond-boson approach 
which we turn to now.

\section{Bond Boson Analysis} \label{sec3}
In this section we detail a mapping of the tetrahedral chain in the
dimerized $S=1$-chain sector onto a system of interacting bosons. To this
end we adapt the well developed bond-operator
method\cite{Sachdev90a,Chubukov91a,gopalan94,Starykh96a,eder98,brenig98}
which has proven to be useful in dimerized spin-$1/2$ systems to the
present situation. We start by introducing a set of singlet-
($s^\dagger_l$), triplet- ($t^\dagger_{l\,\alpha}$), and quintet-bosons
($q^\dagger_{l\,\alpha}$) for each tetrahedron at site $l$. These bosons
create all states within the multiplets ${\cal S}_1$, ${\cal T}_1$, and
${\cal Q}$. The bosons and their corresponding states are listed in table
\ref{tab1}. Note, that we have chosen an $x,y,z$ ($z$) representation for
the triplet (quintet) states. Moreover, the site index is not displayed in
the table.

\begin{table}[tb]
\renewcommand{\arraystretch}{1.1}
\begin{tabular}{l|l|l|l}
 & BB  & $\hat{\alpha}$ & \phantom{I-hate-LaTex}ket \\ \hline
${\cal S}_1$ & $s^\dagger|\rangle$ & 1
&             
$\frac{1}{\sqrt{3}}(|-+\rangle +|+-\rangle -|00\rangle)$
\\ \hline
& $t^\dagger_x|\rangle$ & 2 &
$\frac{1}{2}(|0+\rangle -|+0\rangle +|-0\rangle -|0-\rangle)$
\\
${\cal T}_1$ & $t^\dagger_y|\rangle$
& 3 &
$\frac{i}{2}(|+0\rangle -|0+\rangle +|-0\rangle -|0-\rangle)$
\\
& $t^\dagger_z|\rangle$ & 4 &
$\frac{1}{\sqrt{2}}(|+-\rangle -|-+\rangle)$
\\ \hline
& $q^\dagger_2|\rangle$ & 5 &
$|++\rangle$
\\
& $q^\dagger_1|\rangle$ & 6 &
$\frac{1}{\sqrt{2}}(|+0\rangle+|0+\rangle)$
\\
${\cal Q}$ & $q^\dagger_0|\rangle$
  & 7 &
$\frac{1}{{\sqrt{6}}}(|+-\rangle +2|00\rangle +|-+\rangle)$
\\
& $q^\dagger_{-1}|\rangle$ & 8 &
$\frac{1}{\sqrt{2}}(|-0\rangle+|0-\rangle)$
\\
& $q^\dagger_{-2}|\rangle$ & 9 &
$|--\rangle$
\end{tabular}   
\caption{Bond-boson (BB) representation of the singlet (${\cal S}_1$),
triplet (${\cal T}_1$), and quintet(${\cal Q}$) states in the edge-spin
$S=1$ sector. $|\rangle$ represents the vacuum. $\hat{\alpha}$ refers to
an equivalent running index for each state used to label elements of
$M_{\alpha\hat{\beta}\hat{\gamma}}$ and $N_{\alpha\hat{\beta}\hat{
\gamma}}$ in (\ref{w1}).  Entries in the ket column refer to
$S^z$-eigenstates of $S_{1,2}$ of type $|S^z_1S^z_2\rangle$ with $+,0,-$
denoting $S^z=-1,0,+1$.}
\label{tab1} 
\end{table}

To suppress unphysical states the bosons have to fulfill the usual
hardcore constraint of no double-occupancy
\begin{eqnarray}
s_l^{\dagger}s_l +
t_{l\,\alpha}^\dagger t_{l\,\alpha}^{\phantom{\dagger}} +
q_{l\,\alpha}^\dagger q_{l\,\alpha}^{\phantom{\dagger}} =1
\;\;\;,
\label{w2}
\end{eqnarray}
where doubly appearing Greek indices are to be summed over their
respective ranges. After some straightforward algebra we may express the
$\alpha =x,y,z$ components of the edge-spins $S^\alpha_{l\,1,2}$ by
\begin{eqnarray}
S^\alpha_{l\,1,2}\hat{=}&&
\sqrt{\frac{2}{3}} (\pm s_l^\dagger t_{l\,\alpha}^{\phantom{\dagger}}
\pm t_{l\,\alpha}^\dagger s_l)
-\frac{i}{2}\varepsilon_{\alpha\beta\gamma}
t_{l\,\beta}^\dagger t_{l\,\gamma}^{\phantom{\dagger}}
\nonumber\\
&&\pm M_{\alpha\hat{\beta}\hat{\gamma}}
t_{l\,\hat{\beta}}^\dagger q_{l\,\hat{\gamma}}^{\phantom{\dagger}}
\pm M^\ast_{\alpha\hat{\beta}\hat{\gamma}}
q_{l\,\hat{\gamma}}^\dagger t_{l\,\hat{\beta}}^{\phantom{\dagger}}
+ N_{\alpha\hat{\beta}\hat{\gamma}}
q_{l\,\hat{\beta}}^\dagger q_{l\,\hat{\gamma}}^{\phantom{\dagger}}
\;.\label{w1}
\end{eqnarray}
Since $M_{\alpha\hat{\beta}\hat{\gamma}}$ and
$N_{\alpha\hat{\beta}\hat{\gamma}}$ will remain unused in the remainder of
this work we defer an explicit display of these quantities into appendix
\ref{A}. Inserting (\ref{w1}) into (\ref{w3}) we arrive at the Hamiltonian
\begin{eqnarray}
H_{BB}&=& \frac{a}{2}D+H_0+H_1+H_2+H_3+H_4
\nonumber\\
&& + \sum_l \lambda_l
(s_l^{\dagger}s_l +
t_{l\,\alpha}^\dagger t_{l\,\alpha}^{\phantom{\dagger}} +
q_{l\,\alpha}^\dagger q_{l\,\alpha}^{\phantom{\dagger}}-1)
\label{w4}\\[3pt]
H_0&=& \sum_{l} (
-2 s_{l}^{\dagger}s_{l}^{\phantom{\dagger}}
- t_{l\,\alpha}^{\dagger}t_{l\,\alpha}^{\phantom{\dagger}}
+ q_{l\,\alpha}^{\dagger}q_{l\,\alpha}^{\phantom{\dagger}})
\nonumber \\
H_1&=& -\frac{2 b}{3} \sum_{l} (
t_{l\,\alpha}^{\dagger}
t_{l+1\,\alpha}^{\phantom{\dagger}}
s_{l+1}^{\dagger}
s_{l}^{\phantom{\dagger}} +
t_{l\,\alpha}^{\dagger}
t_{l+1\,\alpha}^{\dagger}
s_{l+1}^{\phantom{\dagger}}
s_{l}^{\phantom{\dagger}} + h.c.)
\nonumber \\
H_2&=& \frac{b}{\sqrt{6}} \sum_{l} (
i \varepsilon_{\alpha\beta\gamma}
t_{l+1\,\alpha}^{\dagger}
t_{l\,\beta}^{\dagger}
t_{l\,\gamma}^{\phantom{\dagger}}
s_{l+1}^{\phantom{\dagger}} + h.c.)
\nonumber \\
H_3&=& -\frac{b}{4} \sum_{l} (
t_{l\,\alpha}^{\dagger}
t_{l+1\,\alpha}^{\dagger}
t_{l+1\,\beta}^{\phantom{\dagger}}
t_{l\,\beta}^{\phantom{\dagger}} -
t_{l\,\alpha}^{\dagger}
t_{l+1\,\beta}^{\dagger}
t_{l+1\,\alpha}^{\phantom{\dagger}}
t_{l\,\beta}^{\phantom{\dagger}})
\nonumber \\
H_4&=& O(q^{(\dagger)})
\nonumber
\;\;\;,
\end{eqnarray}
where $\lambda_l$ is a local Lagrange multiplier to enforce the constraint
(\ref{w2}).  $H_4$ refers to quartic terms involving at least one
quintet and at most one singlet boson. Note that the local Hamiltonian
$H_0$ and the first term $Da/2$ simply reflects the spectrum of
the single tetrahedron.

To treat the interacting bose system (\ref{w4}) approximations have to be
made. To this end we first realize that in the limit of weak
inter-tetrahedral coupling, i.e. $b\ll 1$, the singlet bosons will
condense\cite{Sachdev90a,gopalan94} with $s^{(\dagger)}_l\rightarrow
s\in\Re$.  Focusing on this limit and keeping only terms up to quadratic
order in the boson operators and, moreover, replacing the local Lagrange
multiplier $\lambda_l$ by a global one we arrive at the mean-field theory
(MFT)
\begin{eqnarray}\label{w5}
H_{MFT}&&=
D(-2s^2+\lambda s^2-\lambda+\frac{a}{2})
\nonumber\\
&&
+\sum_{l\,\alpha} (\lambda+1)
q_{l\,\alpha}^\dagger q_{l\,\alpha}^{\phantom{\dagger}}
-\frac{1}{2}\sum_{k\,\alpha}(\lambda-1)
\nonumber \\
+\frac{1}{2}&&\sum_{k\,\alpha}
\Psi_{k\,\alpha}^\dagger
\left[\begin{array}{cc}
\lambda-1+s^2\epsilon_k & s^2\epsilon_k \\
s^2\epsilon_k & \lambda-1+s^2\epsilon_k
\end{array}\right]
\Psi_{k\,\alpha}^{\phantom{\dagger}}
\\ \nonumber \\ \label{7a}
\epsilon_k&&=-\frac{4}{3}b \cos(k)
\;,\end{eqnarray}
where $D$ is the number of dimers and $k$ is a momentum vector.
$\Psi_{k\,\alpha}^{(\dagger)}$ is a a spinor with $\Psi_{k\,\alpha}^\dagger
=[t_{k\,\alpha}^{\dagger}\; t_{-k\,\alpha}^{\phantom{\dagger}}]$ and
$t_{l\,\alpha}^\dagger=\sum_k e^{-ik l}t_{k\,\alpha}^\dagger/\sqrt{D}$. 
The mean-field Hamiltonian can be diagonalized by a Bogoliubov
transformation yielding
\begin{eqnarray}\label{w6}
H_{MFT}&&=
D(\frac{3}{2}-2s^2+\lambda s^2 -\frac{5}{2}\lambda +\frac{a}{2})
\nonumber\\
+\sum_{l\,\alpha} && (\lambda+1)
q_{l\,\alpha}^\dagger q_{l\,\alpha}^{\phantom{\dagger}}
+\sum_{k\,\alpha}
E_k (a_{k\,\alpha}^\dagger a_{k\,\alpha}^{\phantom{\dagger}}+\frac{1}{2})
\;,\end{eqnarray}
\begin{figure}[tb]
\vskip0.2cm
\centerline{\psfig{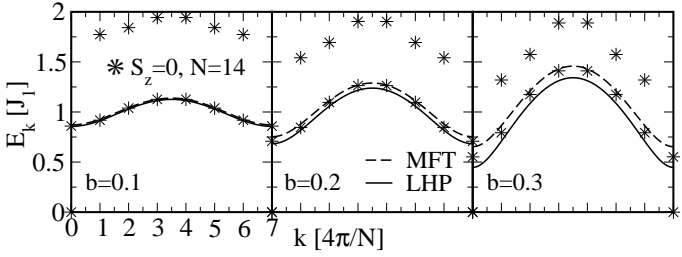}}
\vskip.5cm
\caption{Dashed (solid) line: $E_k$ as from (\ref{w7}) for MFT (LHP).
Stars: first two total-$S_z=0$ excitation of dimerized spin-1 chain
from ED with PBC.}
\label{dispfig}          
\end{figure}
where the threefold degenerate triplet energy $E_k$ is given by
\begin{eqnarray}\label{w7}
E_k = \sqrt{(\lambda-1)^2(1+\frac{s^2}{\lambda-1}2\epsilon_k)}
\end{eqnarray}
and the Bogoliubons $a_{k\,\alpha}^{(\dagger)}$ result from
\begin{eqnarray}\label{w8}
\Psi_{k\,\alpha}^{\phantom{\dagger}}=
\left[\begin{array}{cc}
g_k & h_k \\
h_k & g_k
\end{array}\right]
\Phi_{k\,\alpha}^{\phantom{\dagger}}
\end{eqnarray}
where $\Phi_{k\,\alpha}^{(\dagger)}$ is a a spinor with
$\Phi_{k\,\alpha}^\dagger=[a_{k\,\alpha}^{ \dagger} \;
a_{-k\,\alpha}^{\phantom{\dagger}}]$ and $h^2_k=[(1+\epsilon_k)/E_k-1]/2$,
and $h_k g_k=-\epsilon_k/(2E_k)$ with $h^2_k-g^2_k=1$.  Note that on the
quadratic level the quintet is {\em dispersionless}.
Substituting $d=s^2/(\lambda-1)$ the ground-state energy is\cite{lamdaG0}
\begin{eqnarray}
E^0_{MFT}&&=
D(\frac{3}{2}-2s^2+\lambda s^2 -\frac{5}{2}\lambda +\frac{a}{2})
\nonumber\\
+&&\frac{3}{2} (\lambda-1) \sum_k \sqrt{1+2 d\epsilon_k}
\;\;,
\label{w9}
\end{eqnarray}
where we have used that $\langle t_{l\,\alpha}^\dagger
t_{l\,\alpha}^{\phantom{\dagger}}\rangle=\langle q_{l\,\alpha}^\dagger
q_{l\,\alpha}^{\phantom{\dagger}}\rangle=0$ in the gaped case at $T=0$.
The mean-field order pa\-ra\-me\-ters $s^2$ and $\lambda$ follow from the
saddlepoint conditions $\partial E^0_{MFT}/\partial s^2 = 0$ and $\partial
E^0_{MFT}/\partial\lambda = 0$ which can be combined to result in
\begin{eqnarray}
\frac{5}{2}-d-\frac{3}{2D}
\sum_k\frac{1}{\sqrt{1+2d\epsilon_k}} &=& 0
\label{w10}\\
\lambda-2+\frac{3}{2D}
\sum_k\frac{\epsilon_k}{\sqrt{1+2d\epsilon_k}} &=& 0 
\label{w11}\;\;,
\end{eqnarray}
with (\ref{w10}) independent of $\lambda$. Therefore, only the single
selfconsistency equation (\ref{w10}) has to be solved for $d$ with
$\lambda$ following from direct insertion of $d$ into (\ref{w11}).

In the limit of vanishing inter-tetrahedral coupling, i.e. $b=0$
(\ref{w10},\ref{w11}) reduce to
\begin{eqnarray}
d= 1 \;\;\;,\;\;\; \lambda=2 \;\;\rightarrow \;\; s^2=1
\;.\label{w12}
\end{eqnarray}
This case relates the MFT to the linearized Holstein-Primakoff (LHP)
method\cite{Starykh96a,eder98}, which has found frequent use in bond-boson
approaches to dimerized spin-$1/2$ systems. Within the LHP
the constraint is used to eliminate the singlets on the tetrahedra, i.e.
within $H_0$, by $s_l^{\dagger}s_l= 1- t_{l\,\alpha}^\dagger
t_{l\,\alpha}^{\phantom{\dagger}} - q_{l\,\alpha}^\dagger
q_{l\,\alpha}^{\phantom{\dagger}}$.  Moreover, within $H_{1,...,4}$ the
singlet condensation is implemented with {\em unit} strength, i.e.,
$s^{(\dagger)}_l=1$.  Dropping all terms beyond quadratic order in the
boson operators we arrive at a Hamiltonian which is exactly identical to
(\ref{w5}) with however $\lambda\equiv 2$ and $s^2\equiv 1$.  Therefore
the LHP is {\em identical} to the MFT constrained to (\ref{w12}).  A
priori it is not obvious whether the MFT or the LHP is a more reliable
approximation and we will present results obtained from both methods.

In fig.~\ref{qp2} results for $a_c(b)$ as obtained from (\ref{w9}) are
included for the MFT and LHP approach. At the dimer to singlet-product
phase-boundary the agreement with ED is very good, both for LHP and MFT.
In principle, the singlet condensate restricts the bond-boson approaches
to the dimer phase.  In fact, the LHP spin-gap closes at $b=3/8$ confining
the LHP to $b<3/8<b_c$. The MFT can be continued from the dimer into the
Haldane regime, even though the ground-state symmetries are different,
yielding a transition line qualitatively still comparable to ED.

Next we consider elementary excitations of the dimer state. These may (i)
remain in the dimerized spin-1 chain sector, or (ii) involve transitions
into sectors containing {\em localized} edge-singlets, i.e.~sites with
$T_{i\,l}=0$. In this paper we confine ourselves to the former type. As
has been pointed out in ref.\cite{Johnsson00}, for a single tetrahedron, the
energy of a type-(ii) ${\cal S}_2$-excitation resides within the spin-gap
of the type-(i) excitations for $1/2\leq a\leq 1$. Analogous {\em
dispersionless} singlet gap-states occur in the spin gap of the dimer
phase of the lattice model and will be discussed
elsewhere\cite{WBelsewhere}.  Figure~\ref{dispfig} compares the dispersion
obtained from (\ref{w7}), both for the MFT and LHP for various values of
$b<b_c$, with the first two $S_z=0$ eigenstates obtained from ED on a
finite dimerized spin-1 chain with PBC. Regarding the first triplet
excitations the agreement is very good. A comparison of the spin gap, i.e.
$E_{k=0,\pi}$, as obtained from the MFT and LHP approach in the dimerized
spin-1 chain sector with ED is contained in fig.~\ref{qp3}.  Apart from
the fact that the agreement is satisfactory for $b\lesssim 0.2$ this
figure demonstrates the main difference between the MFT and LHP
approximation. In contrast to the LHP spin-gap which closes for $b>3/8$
the MFT overestimates the binding energy due to dimer formation and
keeps the spin gap opened for all values of $b$.

\section{Two-triplet excitations and Raman scattering}
Raman scattering can be used to probe the total-spin zero excitations of a
spin system at zero momentum. In this section we consider the magnetic
Raman scattering in the dimerized $S=1$ sector of the tetrahedral chain.
Following Loudon and Fleury\cite{Fleury68} the Raman scattering operator
is given by
\begin{eqnarray}
R =&& \sum_{l\,m} a_{lm} 
({\bf E}_i\cdot {\bf n}_{lm})({\bf E}_o\cdot {\bf n}_{lm})
{\bf s}_l\cdot {\bf s}_m
\nonumber \\
= && A E_i E_o \sum_l ({\bf T}_{l\,1}{\bf T}_{l\,2}
+\beta {\bf T}_{l\,2}{\bf T}_{l+1\,1})
\label{wr1}
\;.\end{eqnarray}
${\bf E}_{i(o)}$ are the incoming(outgoing) electric-field vectors and
${\bf n}_{lm}$ are unit vectors connecting exchange-coupled sites.
$a_{lm}$ are matrix element which are identical among each of those
exchange paths corresponding to one of $J_1$, $J_2$, or $J_3$. From this
and the geometry of the tetrahedral chain the second line results for
polarizations of the light along the chain - which we will focus on. While
$\beta$ in (\ref{wr1}) will be of order $b$, it is very unlikely that
$\beta=b$. In the latter case the Raman operator commutes with the
Hamiltonian implying a vanishing Raman intensity at nonzero Raman shifts.
In the former case we use an equivalent Raman operator
\begin{eqnarray}
\tilde{R} = 
R - R|_{\beta=b} =
C \sum_l {\bf T}_{l\,2}{\bf T}_{l+1\,1} 
\label{wr2}
\;,\end{eqnarray}
where $C=A E_i E_o (\beta-b)$. Thus the Raman intensity will be of second
order in $\beta$ and $b$, i.e.~the inter-tetrahedral coupling. To proceed,
we approximate $\tilde{R}$ on the level of the LHP
\begin{eqnarray}
\tilde{R}_{LHP} := &&\lim_{q\rightarrow 0} \tilde{R}_{LHP}(q)
=\lim_{q\rightarrow 0} [-\frac{2C}{3} \sum_k \cos(k+q/2)
\nonumber \\
&&\times (t_{k+q\,\alpha}^{\dagger} +  t_{-k-q\,\alpha}^{\phantom{\dagger}})
(t_{-k\,\alpha}^{\dagger} +  t_{k\,\alpha}^{\phantom{\dagger}})]
\label{wr3}
\;,\end{eqnarray}
where, for later convenience, we have introduced an auxiliary momentum
dependence. The Raman intensity can be obtained from the zero momentum
limit of the dynamical susceptibility
\begin{eqnarray}
\chi(q,\tau)  = \langle T_\tau [\tilde{R}_{LHP}(q,\tau)
\tilde{R}_{LHP}(q,0)]\rangle
\;.\label{wr4}
\end{eqnarray}
Since $\chi(q,\tau)$ is a two-particle propagator, it is important to
assess the relevance of two-particle scattering. In particular, it has
been realized in the context of other dimerized spin-1/2 systems that
magnetic bound states can severely renormalize the bare two-triplet
spectrum\cite{Kotov98,Kotov99,Jurecka00,Trebst00}.

We chose to implement the two-particle scattering within the LHP approach.
Apart from the interactions $H_{2\ldots 4}$ in (\ref{w4}) the constraint
(\ref{w2}) implies a hard-core repulsion between two bosons on a site.  In
the LHP this pertains only to the triplets, as the singlets are condensed
and the quintets have been discarded.  The hard-core is incorporated
directly by introducing an additional contribution to the
Hamiltonian\cite{Kotov98}
\begin{eqnarray}
H_U = U \sum_l
t_{l\,\alpha}^{\dagger}
t_{l\,\beta}^{\dagger}
t_{l\,\alpha}^{\phantom{\dagger}}
t_{l\,\beta}^{\phantom{\dagger}}
\;,\label{wr5}
\end{eqnarray}
with the summation convention on the Greek indices.  $\chi(q,\tau)$ is
evaluated with $H_U$ at finite $U$ and the limit of $U\rightarrow\infty$
is taken at the end.

\begin{figure}[t]
\vskip.2cm
\centerline{\psfig{file=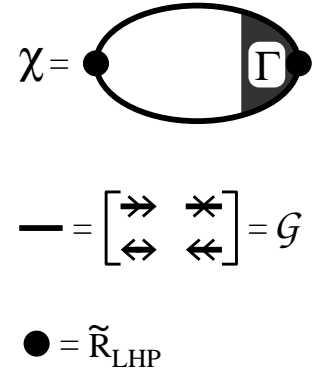,width=4cm}}
\vskip.5cm
\caption{Raman susceptibility. Thick solid lines label the dressed,
2$\times$2 one-triplet, i.e. $t^{(\dagger)}$-particle matrix
Greens\--functions including diagonal and anomalous contributions. The
solid dot refers to the Raman operator (\ref{wr3}). $\Gamma$ is the
two-triplet reducible vertex.}
\label{chiall}
\end{figure}

The Raman susceptibility corresponds to the diagram depicted in
fig.~\ref{chiall}.  To simplify matters we focus on the limit $b\ll 1$.
In that limit the ground state is nearly a pure product of ${\cal S}_1$
singlets and the triplet density induced by quantum fluctuations
$n_t=\langle t_{l\,\alpha}^{\dagger}
t_{l\,\alpha}^{\phantom{\dagger}}\rangle=3\sum_k h^2_k$ is a small
parameter.  As a consequence only the two-triplet {\em
pair}-creation(destruction) vertices contained in (\ref{wr3}) are
relevant. Moreover, contributions to the reducible two-particle propagator
in fig.~\ref{chiall} involving anomalous Greens functions, as well as
one-triplet selfenergy insertions, are suppressed by factors of the triplet
density and will be neglected\cite{Kotov98}. Physically speaking, the
Stokes Raman-process creates two triplets within an approximate singlet
product-state.  These propagate along the tetrahedral chain and form an
interacting two-particle problem with {\em no} additional triplets
generated(destroyed) by quantum fluctuations\cite{Jurecka00}.

\begin{figure}[tb]
\vskip.2cm
\centerline{\psfig{file=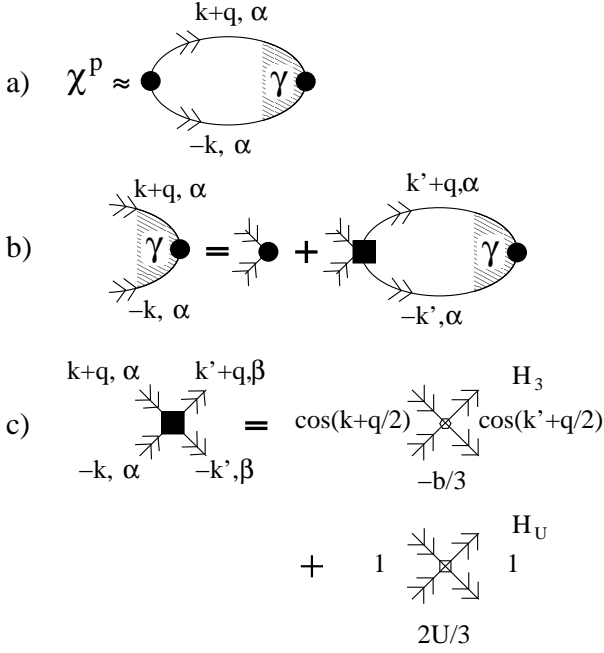,width=8cm}}
\vskip.5cm
\caption{T-matrix approximation to fig. \ref{chiall}: thin,
doubly-directed lines label $11$-elements of the bare one-triplet
Greens function (\ref{wr6}). The solid dot is the two-triplet part
of the Raman vertex (\ref{wr3}). Summation on $k'$ and $\alpha$ is
implied in all bare triplet-bubbles. $\gamma$(solid square) refers
to two-triplet (ir)reducible vertex. Analytic expressions for the
two irreducible vertices due to $H_3$ and $H_U$ are displayed
{\em incorporating} all possible leg exchanges.}
\label{figTmat}
\end{figure}

The two-triplet problem allows for an exact solution {\em including} $H_U$
by the T-matrix approach of fig.~\ref{figTmat}. This figure depicts the
'particle-diagrams', which correspond to the Stokes process. For the
Anti-Stokes process an identical set of 'hole-diagrams' exists with all
lines reversed. In the singlet channel only $H_3$ of (\ref{w4}) and $H_U$
contribute to the irreducible two-particle vertex $\gamma$
(see\cite{triplic_note}).  Due to the momentum space symmetry of $\gamma$
it is convenient to formulate the T-matrix equation using a 2$\times$2
matrix notation.  The bare one-triplet Greens function including normal
and anomalous components is given by
\begin{eqnarray}
G^{ij}_\alpha(k,i\omega_n) && = \frac{1}{(i\omega_n)^2 - E^2_k}
\nonumber\\
&& \times\left[\begin{array}{cc}
i\omega_n +1+\epsilon_k & -\epsilon_k \\
-\epsilon_k & -i\omega_n +1+\epsilon_k
\end{array}\right]
\;,\label{wr6}
\end{eqnarray}
where $\epsilon_k$ and $E_k$ are as of (\ref{7a},\ref{w7},\ref{w12}),
$\alpha=x,y,z$,  and $\omega_n=2n\pi T$.  $G^{ij}_\alpha(k,i\omega_n)$
satisfies the symmetries $G^{11}_\alpha(k,
i\omega_n)=G^{22}_\alpha(k,-i\omega_n)$ and $[G^{21}_\alpha(k,i\omega_n)
]^\ast=G^{21}_\alpha(k,i\omega_n)$. From fig.~\ref{figTmat} we get
\begin{eqnarray}
\chi&&^p(q,z) = 2\chi^{0}_{cc}
\nonumber \\
&& +
2\left[\begin{array}{cc}
\chi^{0}_{cc} \chi^{0}_{c1}
\end{array}\right]
{\bf V}
\left\{
{\bf 1} -
\left[\begin{array}{cc}
\chi^{0}_{cc} & \chi^{0}_{c1} \\
\chi^{0}_{1c} & \chi^{0}_{11}
\end{array}\right]
{\bf V}
\right\}^{-1}
\left[\begin{array}{c}
\chi^{0}_{cc} \\ \chi^{0}_{1c}
\end{array}\right]
,\label{wr7}
\end{eqnarray}
where $\chi^{0}_{cc,c1,1c,11}$ are bare two-particle propagators,
the explicit display of whose momentum and frequency dependence has been
suppressed for brevity. ${\bf V}$ incorporates the momentum independent
coupling-constant factors of the two vertices in $\gamma$ of
fig.~\ref{figTmat}.
\begin{eqnarray}
&&{\bf V} =
\left[\begin{array}{cc}
-b/3 &  \\
& 2U/3
\end{array}\right]
\\[6pt]
&&\chi^{0}_{11}(q,z)  = 3 \sum_k \frac{1}{z-E_{k+q}-E_k}
\stackrel{b\ll 1}{\simeq}
g A 
\label{wr8}\\
&&\chi^{0}_{1c}(q,z)  = 3 \sum_k
\frac{\cos(k+\frac{q}{2})}{z-E_{k+q}-E_k}
\stackrel{b\ll 1}{\simeq}
g (1-\nu A) 
\nonumber \\
&&\chi^{0}_{cc}(q,z) = 3 \sum_k \frac{\cos^2(k+\frac{q}{2})}{z-E_{k+q}-E_k}
\stackrel{b\ll 1}{\simeq}
-g \nu (1-\nu A)
\nonumber 
\;,
\end{eqnarray}
where we have analytically continuated $i\omega_n$ into the upper complex
plane $i\omega_n\rightarrow z$ and have restricted ourselves to the zero
temperature limit. The prefactors of 3 are due to the sum over the triplet
index $\alpha$ and $\chi^{0}_{1c}(q,z) =\chi^{0}_{c1}(q,z)$. The 
two-hole propagator 
$\chi^{h}$ is obtained by reversing the signs of all $E_{(k)k+q}$ in the
denominators of (\ref{wr8}). In the limit $b\ll 1$ one may expand the
square root in (\ref{w7}) which allows for analytic expressions for all of
the $\chi^{0}$'s in terms of the quantities $g$, $\nu$ and $A$
\begin{eqnarray}
&&g=g(q) = \frac{9}{8 b \cos(q/2)} \;\;\;\;\;\;
\nu=\nu(q,z) = \frac{3(z-2)}{8 b \cos(q/2)}
\nonumber \\[6pt]
&&A=A(q,z) = \frac{\mbox{sign}(\mbox{Re}(\nu))}{\sqrt{\nu^2-1}}
\;.\label{wr9}
\end{eqnarray}
From (\ref{wr7}-\ref{wr9}) we obtain the Stokes susceptibility from
$\chi^p(q,z)$ by performing the limit $U\rightarrow\infty$
\begin{eqnarray}\label{wr10}
&&\chi^p(q,z) =
\frac{6[\chi^{0}_{11}\chi^{0}_{cc}-(\chi^{0}_{1c})^2]}
{3\chi^{0}_{11}+b [\chi^{0}_{11}\chi^{0}_{cc}-(\chi^{0}_{1c})^2]} 
\\[12pt]
&&\stackrel{b\ll 1}{\simeq}
\frac{6(\mbox{sign}(\mbox{Re}(\nu))\sqrt{\nu^2-1}-\nu)}
{b[\mbox{sign}(\mbox{Re}(\nu))\sqrt{\nu^2-1}-\nu-8\cos(q/2)/3]\textsl{}}
\label{wr11}\;.
\end{eqnarray}
As in (\ref{wr7}) we refrain from explicitly displaying the momentum and
frequency dependence on the r.h.s.~of (\ref{wr10}). From
(\ref{wr10},\ref{wr11}) one obtains the Raman intensity $I(\omega)$ from
$I(\omega)=-\mbox{Im}\chi^p(0,z\rightarrow\omega+i0^+)$ where $\omega$
refers to the Raman shift.

\begin{figure}[tbh]
\vskip.2cm
\centerline{\psfig{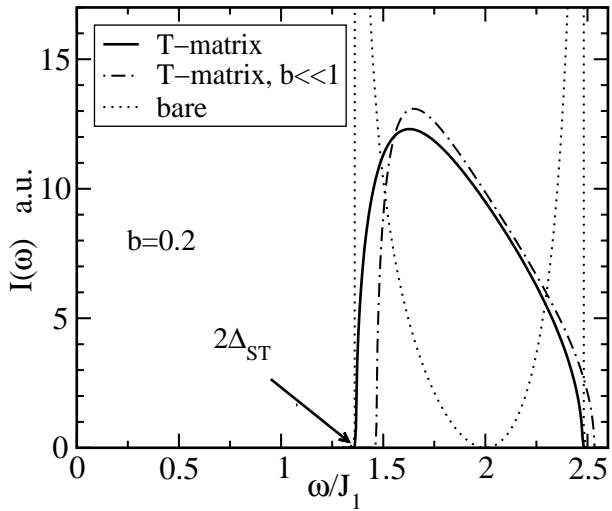}}
\vskip.2cm
\caption{Raman intensity in the dimer phase at $b=0.2$ as obtained
from (\ref{wr10}) (solid), from (\ref{wr11}) dashed-dotted, and from
(\ref{wr7}) for ${\bf V}=0$ (dotted) line. $2\Delta_{ST}$ refers to
twice the singlet-triplet gap.}
\label{ramspec}
\end{figure}

\begin{figure}[bth]
\vskip.2cm
\centerline{\psfig{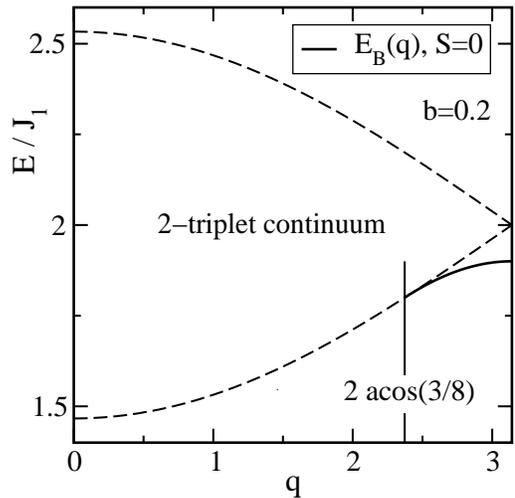}}
\vskip.3cm 
\caption{Two-triplet continuum and dispersion of the $S=0$ bound state
from (\ref{wr11}). Note the y-axis offset.}
\label{BoundState}
\end{figure}

Figure~\ref{ramspec} shows the Raman intensity contrasting the bare
two-triplet spectrum with the interacting one. As is obvious the bare
intensity is strongly renormalized by the two-triplet interactions. In
particular, both of the van-Hove singularities present in the bare
two-triplet spectrum disappear with the almost symmetric shape of the bare
spectrum being deformed by a downward shift of the intensity. These
findings allow for a clear physical interpretation which follows from an
inspection of the denominators of (\ref{wr10},\ref{wr11}). For $q>q_c$
these denominators acquire a zero for energies $E_B(q)$ below the continuum
of the two-triplet scattering states.  I.e., a total-spin zero {\em bound
state} exists in the dimerized spin-one chain at finite
momentum\cite{Ladder01}. Figure~\ref{BoundState} shows the dispersion of
this bound state as obtained from (\ref{wr11}) where
$q_c=2\mbox{acos}(3/8)$.  For $q<q_c$ the bound state turns into a
resonance shortly above the lower edge of the continuum which leads to
the asymmetric Raman intensity of fig.~\ref{ramspec} at $q=0$. This
resonance feature has to be contrasted with the impact of bound states on
the Raman spectra of other dimerized and frustrated low-dimensional
quantum spin systems where $S=0$ collective modes have been observed
rather as sharp excitations {\em within} the spin
gap\cite{lemmens97,lemmens98,lemmens00}.  The actual location of the bound
sate with respect to the two-triplet continuum is significantly affected
by the hard-core repulsion $U$.  Setting $U=0$ in (\ref{wr7}) the
short-range attraction due to $H_3$ would be overestimated with $E_B(q)$
resulting from $1+b\chi^{0}_{cc}(q,E_B(q))/3=0$ which would yield a bound
state {\em below} the lower edge of the continuum for all $q$.  While
$E_B(q)$ in fig.~\ref{BoundState} has been plotted in units of $J_1$ for
$b=0.2$ all bound state dispersions can be {\em rescaled} onto a single
one in terms of the frequency variable $\nu$. This is certainly an
artifact of the limit $b\ll 1$.  Finally we note that the relative
agreement between (\ref{wr10}) and (\ref{wr11}) improves continuously as
$b\rightarrow 0$.

\section{Conclusion}
In summary we have investigated the ground state and several aspects of
the one- and two-triplet excitations of a tetrahedral cluster spin-chain.
A number of open questions remain. In particular excitations involving
localized edge singlets of the tetrahedra are an issue yet to be resolved.
In the case that such excitations are Raman active we expect them to lead
to a dispersionless distribution of intensities which can reside in the
spin-gap of fig.~\ref{ramspec} for certain ranges of the parameters
$(a,b)$. Below a temperature of $T\lesssim 50$K the Raman intensity on
Cu$_2$Te$_2$O$_5$Br$_2$ gradually builds up a continuum\cite{Lemmens01z}
centered at 60cm$^{-1}$ which, below $T\lesssim 8$K is accompanied by an
additional sharp mode developing at 20cm$^{-1}$. One might speculate the
continuum to correspond to that of fig.~\ref{ramspec} and the sharp mode
to consist of transitions involving edge-singlets. Yet, the measured
continuum is rather more symmetric than the solid line in
fig.~\ref{ramspec}. This might be related to the effects of three-dimensional
couplings between the tetrahedra in the tellurates leaving their magnetism
an open issue which deserves further studies. Finally the role of
perturbations breaking the complete frustration may be of relevance in the
vicinity of the first-order transition leading to additional quantum
phases.

{\bf Acknowledgments:}
It is a pleasure to thank P. Lemmens, R. Valenti, C. Gros, F. Mila, E.
Kaul, and Ch. Geibel for stimulating discussions and comments.  This
research was supported in part by the Deutsche Forschungsgemeinschaft
under Grant No. BR 1084/1-1 and BR 1084/1-2 and trough SFB 463.

\begin{appendix}
\section{${\cal T}_1 \leftrightarrow {\cal Q}$ and
${\cal Q} \leftrightarrow {\cal Q}$
transition matrix elements}\label{A}

In table (\ref{tab2},\ref{tab3}) of this appendix we list the nonzero
matrix elements of $M_{\alpha\hat{\beta}\hat{\gamma}}$ and
$N_{\alpha\hat{\beta}\hat{\gamma}}$ from (\ref{w1}).
One should note, that there are no transitions mediated by
$S^\alpha_{l\,1,2}$ between ${\cal S}_1$ and ${\cal Q}$.
\begin{table}[h]
\begin{tabular}[c]{l|l|l}
$\hat{\beta}\;\;\;\hat{\gamma}\;\;\;2^{3/2}M_{x\hat{\beta}\hat{\gamma}}$
&                                                  
$\hat{\beta}\;\;\;\hat{\gamma}\;\;\;2^{3/2}M_{y\hat{\beta}\hat{\gamma}}$
&
$\hat{\beta}\;\;\;\hat{\gamma}\;\;\;2^{3/2}M_{z\hat{\beta}\hat{\gamma}}$
\\ \hline &&\\[-8pt]
$\matrix{ 2 & 5 & 1 \cr 2 & 7 & -{\sqrt{\frac{2}
        {3}}} \cr 2 & 9 & 1 \cr 3 & 5 & \imag  \cr 3 & 9 & -\imag
      \cr 4 & 6 & -1 \cr 4 & 8 & 1 \cr  }$
&
$\matrix{ 2 & 5 & \imag  \cr 2 & 9 & -\imag
      \cr 3 & 5 & -1 \cr 3 & 7 & -{\sqrt{\frac{2}{3}}} \cr 3 & 9 & 
     -1 \cr 4 & 6 & -\imag  \cr 4 & 8 & -\imag  \cr  }$
&
$\matrix{ 2 & 6 & 
     -1 \cr 2 & 8 & 1 \cr 3 & 6 & -\imag  \cr 3 & 8 & -\imag  \cr
     4 & 7 & 2\,{\sqrt{\frac{2}{3}}} \cr  }$
\end{tabular}   
\caption{Nonzero transition-operator matrix-elements, see (\ref{w1}),
conecting ${\cal T}_1$ and ${\cal Q}$ as of table \ref{tab1}.}
\label{tab2}
\end{table}
\begin{table}[h]
\renewcommand{\arraystretch}{2}
\begin{tabular}[c]{l|l|l}
$\hat{\beta}\;\;\;\hat{\gamma}\;\;\;2 N_{x\hat{\beta}\hat{\gamma}}$
&                                                  
$\hat{\beta}\;\;\;\hat{\gamma}\;\;\;2 N_{y\hat{\beta}\hat{\gamma}}$
&
$\hat{\beta}\;\;\;\hat{\gamma}\;\;\;2 N_{z\hat{\beta}\hat{\gamma}}$
\\ \hline &&\\[-20pt]
$\matrix{ 5 & 6 & 1 \cr 6 & 5 & 1 \cr 6 & 7 & {\sqrt{\frac{3}
       {2}}} \cr 7 & 6 & {\sqrt{\frac{3}{2}}} \cr 7 & 8 & {\sqrt{\frac{3}
       {2}}} \cr 8 & 7 & {\sqrt{\frac{3}
       {2}}} \cr 8 & 9 & 1 \cr 9 & 8 & 1 \cr  }$
&
$\matrix{ 5 & 6 & -\imag  \cr 6 & 5 & \imag
     \cr 6 & 7 & -\imag \,{\sqrt{\frac{3}{2}}} \cr 7 & 6 & \imag \,
    {\sqrt{\frac{3}{2}}} \cr 7 & 8 & -\imag \,
    {\sqrt{\frac{3}{2}}} \cr 8 & 7 & \imag \,
    {\sqrt{\frac{3}{2}}} \cr 8 & 9 & -\imag  \cr 9 & 8 & \imag  \cr  }$
&
$\matrix{ 5 & 5 & 2 \cr 6 & 6 & 1 \cr 8 & 8 & 
     -1 \cr 9 & 9 & -2 \cr  }$
\end{tabular}   
\caption{Nonzero transition-operator matrix-elements, see (\ref{w1}),
within ${\cal Q}$ as of table \ref{tab1}.}
\label{tab3}
\end{table}
\end{appendix}


\begin{thebibliography}{99}
\bibitem{triplic_note}
In principle two-particle irreducible vertices can be constructed to
second order in $b$ from $H_2$. In the singlet channel, however, because
of the symmetries of the spin indices in (\ref{wr3}) w.r.t. those of
the Levi-Civita symbol in $H_2$ these vanish identically.
\end{thebibliography}
\end{document}